\documentclass[article,a4paper,bm,twocolumn,showpacs,superscriptaddress]{revtex4}
\usepackage{bm}
\usepackage{amssymb}
\usepackage{amsfonts}
\usepackage{amsmath}
\usepackage{graphicx}
\usepackage[dvips]{color} 

\definecolor{g-blue}{rgb}{0.83,0.95,1}
\definecolor{Blue}{rgb}{0.5,0.5,1}
\definecolor{DarkBlue}{rgb}{0.00,0.00,0.58}
\definecolor{g-yellow}{rgb}{1,1,0.7}
\definecolor{g-green}{rgb}{0.9,1,0.9}
\definecolor{green}{rgb}{0,0.6,0}
\definecolor{Green}{rgb}{0,0.4,0}
\definecolor{cyan}{rgb}{0,0.7,0.7}
\definecolor{black}{rgb}{0,0,0}
\definecolor{grey}{rgb}{0.4 ,0.4 ,0.4 }

\begin{document}

\def\Fbox#1{\vskip1ex\hbox to 8.5cm{\hfil\fboxsep0.3cm\fbox{%
  \parbox{8.0cm}{#1}}\hfil}\vskip1ex\noindent}  


\newcommand{\eq}[1]{(\ref{#1})}
\newcommand{\Eq}[1]{Eq.~(\ref{#1})}
\newcommand{\Eqs}[1]{Eqs.~(\ref{#1})}
\newcommand{\Fig}[1]{Fig.~\ref{#1}}
\newcommand{\Figs}[1]{Figs.~\ref{#1}}
\newcommand{\Sec}[1]{Sec.~\ref{#1}}
\newcommand{\Secs}[1]{Secs.~\ref{#1}}
\newcommand{\Ref}[1]{Ref.~\cite{#1}}
\newcommand{\Refs}[1]{Refs.~\cite{#1}}

\def\be{\begin{equation}}\def\ee{\end{equation}}
\def\bea{\begin{eqnarray}}\def\eea{\end{eqnarray}}
\def\bse{\begin{subequations}}\def\ese{\end{subequations}}
\newcommand{\BE}[1]{\begin{equation}\label{#1}}
\newcommand{\BEA}[1]{\begin{eqnarray}\label{#1}}
\newcommand{\BSE}[1]{\begin{subequations}\label{#1}}

\let \nn  \nonumber  \newcommand{\br}{\\ \nn}
\newcommand{\BR}[1]{\\ \label{#1}}
\def\hf{\frac{1}{2}}
\let \= \equiv \let\*\cdot \let\~\widetilde \let\^\widehat \let\-\overline
\let\p\partial \def\pp {\perp} \def\pl {\parallel}
\def\ort#1{\^{\bf{#1}}}
\def\Trans{^{\scriptscriptstyle{\rm T}}}
\def\x{\ort x} \def\y{\ort y} \def\z{\ort z}
 \def\bn{\bm\nabla} \def\1{\bm1} \def\Tr{{\rm Tr}}
\def\Re{\mbox{  Re}}
\def\<{\left\langle}    \def\>{\right\rangle}
\def\({\left(}          \def\){\right)}
 \def \[ {\left [} \def \] {\right ]}

\renewcommand{\a}{\alpha}\renewcommand{\b}{\beta}\newcommand{\g}{\gamma}
\newcommand{\G} {\Gamma}\renewcommand{\d}{\delta}
\newcommand{\D}{\Delta}\newcommand{\e}{\epsilon}\newcommand{\ve}{\varepsilon}
\newcommand{\E}{\Epsilon}\renewcommand{\o}{\omega} \renewcommand{\O}{\Omega}
\renewcommand{\L}{\Lambda}\renewcommand{\l}{\lambda}
\renewcommand{\t}{\tau}
\def\r{\rho}\def\k{\kappa}
\def\t{\theta } \def\T{\Theta } \def\s{\sigma} \def\S{\Sigma}

\newcommand{\B}[1]{{\bm{#1}}}
\newcommand{\C}[1]{{\mathcal{#1}}}    
\newcommand{\BC}[1]{\bm{\mathcal{#1}}}
\newcommand{\F}[1]{{\mathfrak{#1}}}
\newcommand{\BF}[1]{{\bm{\F {#1}}}}

\renewcommand{\sb}[1]{_{\text {#1}}}  
\renewcommand{\sp}[1]{^{\text {#1}}}  
\newcommand{\Sp}[1]{^{^{\text {#1}}}} 
\def\Sb#1{_{\scriptscriptstyle\rm{#1}}}

\title{Andreev reflection in rotating superfluid $^3$He-B}

\author{V.B.~Eltsov}
\affiliation{O.V. Lounasmaa Laboratory, School of Science, Aalto University, POB 15100, FI-00076 AALTO, Finland}

\author{J.J.~Hosio$^\ddagger$}
\affiliation{O.V. Lounasmaa Laboratory, School of Science, Aalto University, POB 15100, FI-00076 AALTO, Finland}


\author{M.~Krusius}
\affiliation{O.V. Lounasmaa Laboratory, School of Science, Aalto University, POB 15100, FI-00076 AALTO, Finland}

\author{J.T. M\"akinen}
\affiliation{O.V. Lounasmaa Laboratory, School of Science, Aalto University, POB 15100, FI-00076 AALTO, Finland}

\date{\today}

\begin{abstract}

Andreev reflection of quasiparticle excitations from quantized line vortices is reviewed in the isotropic B phase of superfluid $^3$He in the temperature regime of ballistic quasiparticle transport at $T \leq 0.20\,T_\mathrm{c}$. The reflection from an array of rectilinear vortices in solid-body rotation is measured with a quasiparticle beam illuminating the array mainly in the orientation along the rotation axis. The result is in agreement with the calculated Andreev reflection. The Andreev signal is also used to analyze the spin down of the superfluid component after a sudden impulsive stop of rotation from an equilibrium vortex state. In a measuring setup where the rotating cylinder has a rough bottom surface, annihilation of the vortices proceeds via a leading rapid turbulent burst followed by a trailing slow laminar decay from which the mutual friction dissipation can be determined. In contrast to currently accepted theory, it is found to have a finite value in the zero temperature limit: $\alpha (T \rightarrow 0) = (5 \pm 0.5) \cdot 10^{-4}$.

\end{abstract}
%

\maketitle 

\section{Introduction}
\label{Intro}

Andreev reflection \cite{Andreev-1}, the celebrated phenomenon which Alexander Andreev introduced in 1964 to explain the increased resistance in heat flow through a normal metal -- superconductor interface, took three decades to be demonstrated in superfluid $^3$He-B \cite{He3Andreev-1}. Nevertheless, during the more recent past it has become one of the prime tools to study the zero temperature limit, $T \rightarrow 0$, of this charge-neutral p-wave fermion system. If the temperature is sufficiently low, so that collisions between quasiparticle excitations are practically absent in the bulk volume and ballistic propagation prevails, then measurements with vibrating sensors typically prominently display characteristic signatures from Andreev reflection.

Rotation is another central research tool of superfluid $^3$He. During the past decade it has been applied at ever lower temperatures so that Andreev reflection measurements have become possible even in rotating flow. Research on rotating $^3$He superfluids  has been centered in Helsinki since the early 1980ies and has been vital for exploring the many different forms of quantized vorticity, both the structure as well as the dynamics. Alexander Andreev was one of the original founders of this research effort.

A well-known example of Andreev reflection, originally demonstrated with vibrating wire resonators by the Lancaster group in 2001 \cite{He3Andreev-2}, is shown in Fig.~\ref{AndreevDemo}. Here two mechanical oscillators in close proximity to each other are vibrating in a bath of superfluid $^3$He-B. One of them is driven at high displacement amplitude such that its strong vibrations generate a turbulent tangle of quantized vortices which shrouds both vibrators. The second vibrator is driven at low amplitude such that its output measures the damping of its oscillations by the  thermal (but ballistic) quasiparticles. In spite of the heat input by the generator and the resulting increased overall number of excitations, the detector, shielded by the cloud of tangled vortices, displays reduced damping: it is hit by fewer quasiparticles from the surrounding cloud of thermal excitations and thus records a lower apparent temperature. This counterintuitive result is caused by the Andreev retroreflection shadow cast by the vortex tangle.

\begin{figure}[t!!]
\begin{center}
\centerline{\includegraphics[width=0.9\linewidth]{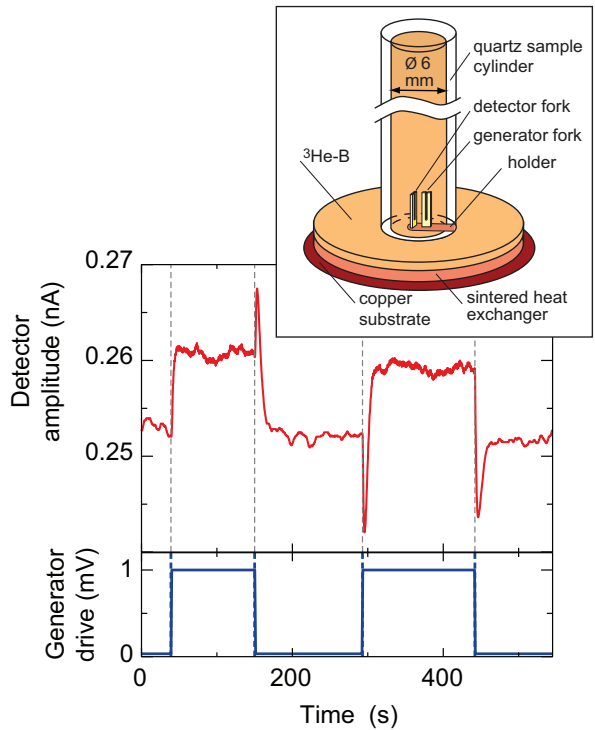}}
\caption{Demonstration of Andreev reflection \cite{Forks-2}. As sketched in the inset above, two quartz tuning fork oscillators are vibrating in a container filled with $^3$He-B at 0.20\,\textit{T}$_\mathrm{c}$. A drive with square wave envelope at high amplitude is fed to one of the forks which generates a turbulent vortex tangle. The tangle spreads around the two forks and casts an Andreev shadow which is recorded as reduced damping by the second sensor fork driven at a one order of magnitude lower excitation amplitude.
}
\label{AndreevDemo}
\end{center}
\vspace{-10mm}
\end{figure}

The example in Fig.~\ref{AndreevDemo} \cite{Forks-2} demonstrates the possibilities of using Andreev reflection for the study of a clean Fermi system in the very low temperature limit. This work and the development of the appropriate techniques has long been the domain of the Lancaster group \cite{Lancs-1}. More recently the ultra-low temperature regime of $^3$He-B has moved in the forefront of general interest when it was realized that conventional sub-mK refrigeration techniques by means of adiabatic demagnetization cooling of copper are quite adequate if the total heat leak to the sample volume can be reduced to below $\sim 20\,$pW. This recognition has led to the development of new measuring techniques for the ballistic regime in superfluid $^3$He which are based on the use of mechanical vibrating resonance devices \cite{Camera} or the existence of a novel coherently precessing NMR mode \cite{MagnonBEC-1}. Such work has been driven by the hope to reveal explicit new information on Andreev reflection, or on the existence of Andreev bound states on surfaces, interfaces, and vortex cores, or the expectation to identify the Majorana character in the spectrum of the bound state excitations \cite{MajoranaReview}.

Today the experimental tools for making use of Andreev reflection consist of a quasiparticle radiator, a box with a heating element coupled to the $^3$He-B bath via a small orifice which defines the beam, and a sensor in the bath, which traditionally has been a highly sensitive vibrating wire resonator. More recently the mass-produced quartz tuning fork oscillator \cite{Forks-1} has been found to have sufficient sensitivity as a quasiparticle detector and because of its easier use and insensitivity to magnetic fields it has gained in popularity.

The theoretical basis of Andreev reflection in superfluid $^3$He, with emphasis on the phenomena arising from the spin-triplet and orbital p-wave pairing, was laid out in an early review of Kurkij\"arvi and Rainer \cite{Kurki&Rainer}. The present brief overview discusses measurements on Andreev reflection in the rotating B phase, describing how Andreev reflection can be used to study quantized vortex lines and their dynamics. This is an obvious area where Andreev reflection measurements have great potential, especially concerning quantum turbulence, the peculiar characteristic of superfluid flow in the limit of weak mutual friction damping, which has been in the focus of recent interest \cite{Lancs-1}.

Originally Andreev reflection measurements in rotation became necessary as a means to calibrate reflection from quantized line vortices \cite{He3Andreev-3}. This provides the starting point for the discussion below. The same experimental setup can also be used to record the dynamic response of a rotating vortex array. The most common type of such measurements is the determination of the response to a sudden stop of the rotation drive. Based on our knowledge of superfluid flows in $^4$He, it has been thought that laminar flow of quantized line vortices becomes unstable in the limit of vanishing mutual friction: when $\alpha (T) \rightarrow 0$ any minute perturbations in the flow can lead to instabilities and cause a tangle of vortices to be formed, so that new dissipation mechanisms become available, primarily fueled by reconnections between neighbouring vortices. Andreev reflection provides one of the means to monitor the flow of vortices. Recent measurements on the rate of vortex annihilation after a sudden reduction of the rotation velocity have shown that in $^3$He-B turbulence is not necessarily the only form of response \cite{Hosio3Heflow}. Depending on the geometry, surface properties, etc., one finds that the rate of vortex decay in cylindrically symmetric flow may correspond to the slow mutual friction damping of laminar flow or to a faster process brought about by additional turbulent dissipation mechanisms. Some of these studies will be discussed below.

\section{Andreev shadow of a vortex}
\label{AndreevShadow}

The theory of Andreev reflection is particularly straightforward in the ballistic temperature regime, as has been demonstrated by Barenghi et al. \cite{AndreevTheory}, who using Hamiltonian mechanics calculate the trajectory of a quasiparticle scattered by a single rectilinear vortex (Fig.~\ref{AndreevScatter}).

\begin{figure}[t!!]
\begin{center}
\centerline{\includegraphics[width=0.8\linewidth]{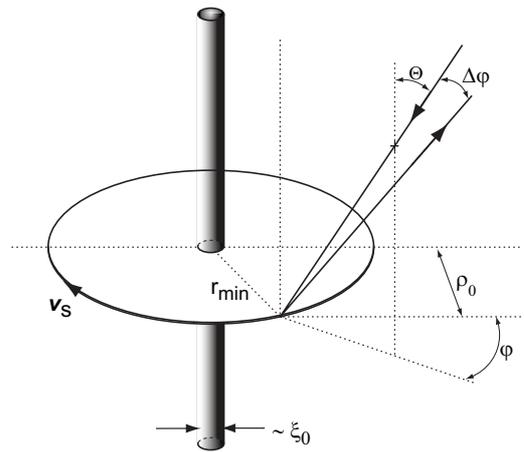}}
\caption{Trajectory of a particle excitation which undergoes Andreev reflection on approaching a rectilinear quantized line vortex and then retraces its path as a hole excitation with a very small deflection by $\Delta \varphi$ .}
\label{AndreevScatter}
\end{center}
\vspace{-10mm}
\end{figure}

The quasiparticle moves with the kinetic energy
\begin{equation}
\label{qp-energy-1}
\epsilon_\mathrm{p} = \frac {p^2}{2 m^*} - \epsilon_\mathrm{F} \;
\end{equation}
with respect to the Fermi energy $\epsilon_\mathrm{F} = p_\mathrm{F}^2/(2 m^*)$, where $\mathbf{p}$ is the linear quasiparticle momentum and $m^*$ its effective mass (we refer here to the 29\,bar pressure of the $^3$He liquid in the measurements of Sec.~\ref{Results} with $m^* \approx 5.4\, m$, where $m$ is the bare mass of the $^3$He atom). The superfluid circulation $\kappa = h/(2m)$ trapped around the vortex core corresponds to an azimuthally circulating superfluid flow with the velocity $\mathbf{v}_\mathrm{s} = \frac {\kappa} {2\pi r} \, \hat{\mathbf{e}}_\varphi$. In this flow field the energy of the quasiparticle is transformed to
\begin{equation}
\label{qp-energy-2}
E = \sqrt{\epsilon_\mathrm{p}^2 + \Delta_0^2} + \mathbf{p} \cdot \mathbf{v}_\mathrm{s}  \; ,
\end{equation}
where $\Delta (T \rightarrow 0) = \Delta_0$ is the energy gap in the zero temperature limit ($k_\mathrm{B} T < 0.1\,\Delta_0$). The trajectory of the particle can be traced from the Hamiltonian equations of motion $d\mathbf{r}/dt =   \partial E/\partial \mathbf{p}$ ($=\mathbf{v}_\mathrm{g}$, the group velocity of the excitations) and $d\mathbf{p}/dt = - \partial E/\partial \mathbf{r}$. Two length scales are involved: namely (1) describing the change in the order parameter amplitude at the vortex, the vortex core radius, measured in terms of the superfluid coherence length $\xi_0 = \hbar v_\mathrm{F} / (\pi \Delta_0) \sim 10\,$nm, while (2) the scattering process is characterized by an angular momentum $p_\varphi \rho_0$ (where $p_\varphi = \mathbf{p} \cdot \hat{\mathbf{e}}_\varphi$), which is a constant of the motion and defines the second length scale, the impact parameter $\rho_0$.

In the ballistic regime the quasiparticle energy spectrum resembles the roton minimum in superfluid $^4$He, since
\begin{equation}
\sqrt{\epsilon_\mathrm{p}^2 + \Delta_0^2} \approx \Delta_0 + \frac {(p-p_\mathrm{F})^2}{2 \Delta_0 v_\mathrm{F}^2}\,.
\end{equation}
An incoming quasiparticle with its energy above the minimum in the range
\begin{equation}
E > \Delta_0 + \frac {\kappa} {2 \pi \rho_0} \, p_\mathrm{F}  \hspace {5mm} \rm{[no \; reflection]}
\end{equation}
is found to follow a usual straight trajectory past the vortex retaining its particle nature. For particle excitations this is the case on that side of the vortex where $\mathbf{p} \cdot \mathbf{v}_\mathrm{s} =  \frac {\kappa} {2 \pi \rho_0} \, p_\varphi < 0$, while on the opposite side there are no allowed states for a particle in the range
\begin{equation}
\Delta_0 < E <  \Delta_0 + \frac {\kappa} {2 \pi \rho_0} \, p_\mathrm{F}  \hspace{6mm} \rm{[Andreev \; reflected]}
\end{equation}
and it is found to retroreflect \cite{Pickett}, \textit{ie.} the particle - hole symmetry is broken and the excitation changes character (and sign) from a particle to a hole. It also almost retraces its trajectory of incidence, being deflected only by a very small angle
\begin{equation}
\label{AndreevAngle}
\Delta \varphi = \frac {\hbar} {p_\mathrm{F}} \, \sqrt{\frac {\pi} {5 \xi_0 \rho_0}} \ll 1\,.
\end{equation}
The distance of closest approach to the vortex core, where the retroreflection occurs, is found to be
\begin{equation}
\label{r-min}
r_\mathrm{min} \approx \sqrt{5\pi \xi_0 \rho_0} \; \frac   {\Delta_0} {\epsilon_\mathrm{p}} \sim 10\,\mu\mathrm{m}
\end{equation}
and is thus inversely proportional to the kinetic energy $\epsilon_\mathrm{p}$ of the incoming particle while the length scale is the geometric mean of $\xi_0$ and $\rho_0$. The upper limit for the impact parameter, where retroreflection can still occur, is
\begin{equation}
\label{r-c}
\rho_{0 \mathrm{c}} \approx 5\pi \xi_0  \left( {\frac   {\Delta_0} {\epsilon_\mathrm{p}}}\right)^2 \,.
\end{equation}
Its magnitude is of order $\sim 10^3 \xi_0$ for thermal quasiparticles (with $\epsilon_\mathrm{p} \sim k_\mathrm{B} T$) and thus the Andreev shadow of the vortex becomes experimentally significant, even though it is restricted to only one side of the vortex for one species of excitations. In practice rotating measurements proceed in the low-density limit where the Andreev shadows of neighbouring vortices can be considered approximately additive. For denser 3-dimensional turbulent vortex structures the problem of tracing the excitation trajectories becomes complex \cite{NumCalc-1,NumCalc-2}, the total Andreev shadow cannot be estimated from the contribution of single vortices, and one has to resort to numerical Monte-Carlo-type simulation calculations.

\section{Experimental principles}
\label{ExpPrinciples}

For quantitative experimental measurements a controlled setup with an oriented beam of quasiparticles and a well-known distribution of vortices is required. In the experiment of Fig.~\ref{AndreevDemo}, for instance, the usual assumption is that the vortex tangle is homogeneous and isotropic in all directions \cite{TurbDecay}. This is a simplification, since the prongs of the tuning fork oscillator vibrate in antiphase in the plane in which they are contained. When the fork excitation is increased above a critical value, vortex rings are generated and shed off from the prongs. The rings propagate with their self-velocity $v_\mathrm{ring} \propto 1/R_\mathrm{ring}$, and because of the spread in their size distribution, they ultimately collide forming a tangle via reconnections \cite{RingCalc}. Owing to the oriented motion of the prongs, the ejection pattern of the rings is not isotropic and thus the orientational homogeneity or the spatial extent of the tangle are not expected to be uniform \cite{Tsubota}. As Andreev reflection depends on both the configuration and density of vortices with respect to the incident quasiparticle beam, a better controlled measurement is needed for calibration purposes. This can be achieved using the rotating equilibrium state in a cylindrical container (Fig.~\ref{ExpScheme}). Here the configuration is fixed and the number of vortices $N_\mathrm{v} \approx \pi R^2 \, n_\mathrm{v}$ can be externally adjusted by manipulating the angular rotation velocity $\Omega$, which controls the aerial density $n_\mathrm{v} = 2\Omega/\kappa$ and thus the inter-vortex distance $\ell \sim 1/\sqrt{n_\mathrm{v}}$ of the rectilinear line vortices in the rotating cylinder of radius $R$.

\begin{figure}[t!]
\begin{center}
\centerline{\includegraphics[width=0.8\linewidth]{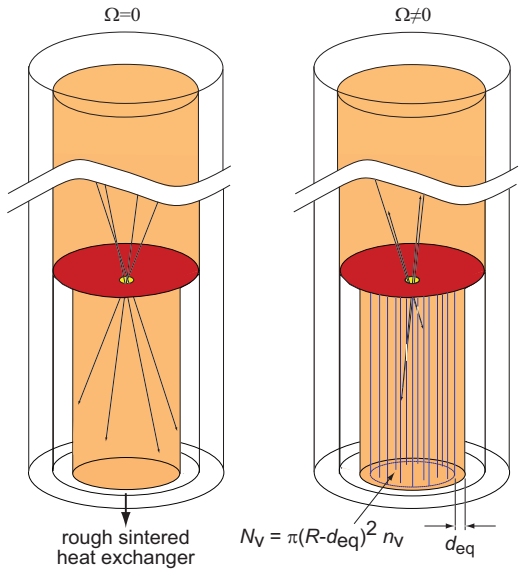}}
\caption{Principle of the rotating calibration experiment for Andreev reflection. At rest ($\Omega$ = 0) on the left, all excitations which are not scattered back through the orifice (owing to diffuse scattering processes on the walls) are thermalized in the heat exchanger at the bottom. In rotation ($\Omega \neq 0$) on the right, part of the beam is Andreev reflected from the equilibrium vortex state below the orifice and the density of excitations above the orifice is increased.}
\label{ExpScheme}
\end{center}
\vspace{-10mm}
\end{figure}

In rotation metastability in the total number of vortices can arise from the presence of a non-negligible critical angular velocity increment $\Omega_\mathrm{c}$. It determines the flow velocity at the cylindrical wall at which a vortex is formed, while rotation is increased: $\mid \mathbf{v}_\mathrm{n} - \mathbf{v}_\mathrm{s} \mid = \Omega_\mathrm{c} R$ \cite{VorForm}. This process controls the total number of vortices in increasing rotation, while the density is fixed to the equilibrium rotation value $n_\mathrm{v} = 2\Omega/\kappa$. In contrast, annihilation of vortices on the cylindrical container wall is not associated with any appreciable energy barrier \cite{Annihilation}. Thus the reference state is preferably formed making use of the threshold to annihilation and not of vortex formation, which means that our rotating reference state is the equilibrium vortex state where the centrally located vortex cluster is surrounded and separated from the cylindrical wall by a vortex-free confining annulus of width $d_\mathrm{eq}$ which in practice is usually the minimum possible width \cite{Annihilation} and only slightly larger than the inter-vortex distance: $d_\mathrm{eq} \gtrsim 1/\sqrt{n_\mathrm{v}}$. So far in the very low temperature measurements the inter-vortex distance is much larger, with $1/\sqrt{n_\mathrm{v}} \gtrsim 0.1$\,mm, than the radius of the Andreev shadow [Eq.~(\ref{r-c})].

The recipe to create the equilibrium vortex state with $N_\mathrm{eq} (\Omega)$ vortices is to increase rotation well above the target velocity and then to decrease $\Omega$ so that vortices have annihilated when the desired value of $\Omega$ is reached. These operations are preferably performed at high mutual friction above $0.7\,T_\mathrm{c}$ and subsequently the sample is cooled at constant rotation to the desired temperature. This precaution becomes necessary since after annihilation the remaining vortex array relaxes to its equilibrium configuration which at the lowest mutual friction values is a slow process requiring hours.

\begin{figure}[t!]
\begin{center}
\centerline{\includegraphics[width=0.8\linewidth]{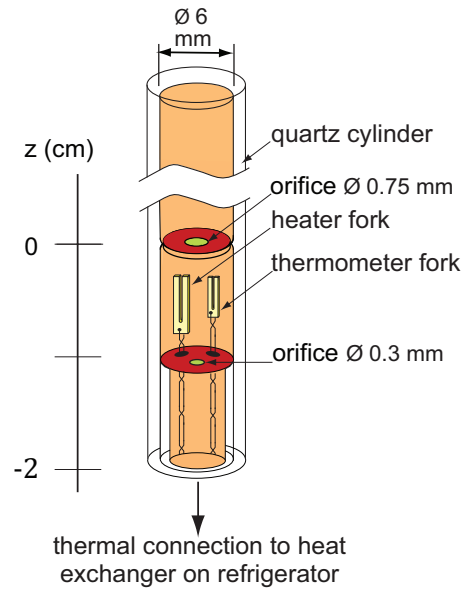}}
\caption{Experimental setup for calibrating Andreev reflection from a rotating array of quantized vortices. Heat flow is mainly vertical directed towards the heat exchanger on the bottom. The volume above the 0.3\,mm orifice functions as a bolometer for measuring the density of excitations when heated with the heater fork. The upper division plate with an 0.75~mm diameter aperture blocks the flow of vortices into the topmost compartment from below, where the critical rotation velocity of vortex formation is lower.  }
\label{Setup}
\end{center}
\vspace{-10mm}
\end{figure}

In the setup of Fig.~\ref{Setup}, the rotating cylinder is compartmentalized with two division walls in three sections. The lower division wall with a small orifice of 0.3\,mm diameter is the main thermal resistance along the cylindrical tower. It is used to measure the change in the quasiparticle density above the orifice as a function of $N_\mathrm{eq} (\Omega)$. The upper large orifice of 0.75\,mm diameter restricts the flow of vortices from the middle section to the uppermost compartment. In the uppermost section the critical velocity is defined by the smooth fused quartz walls of the cylinder so that it is characterized by a high value $\Omega_\mathrm{c} \gtrsim 1.5 \, $ rad/s. It can be rotated in the metastable vortex-free state, the so-called Landau state, up to $\Omega < \Omega_\mathrm{c}$. In contrast, the lowermost section has a rough bottom of sintered copper powder which reduces its critical velocity to $\Omega_\mathrm{c} \sim 0.1 \, $ rad/s. Likewise, the middle section with the two quartz tuning fork oscillators and their leads has a reduced $\Omega_\mathrm{c}$. Thus after rotational deceleration, when vortices have been annihilated, one will find the equilibrium vortex state in the two lower sections, while in the uppermost section the vortex number can be adjusted to have any value between $0 \leq N_\mathrm{v} \leq N_\mathrm{eq}$ \cite{VorForm}.

In Fig.~\ref{Setup}, the heat flow carried by quasiparticle excitations moves along the tower towards the sintered heat exchanger which is always the coldest place, the heat sink. The dominant thermal resistance which it encounters is the orifice of 0.3\,mm diameter in the lower division plate. It defines the quasiparticle beam for the Andreev reflection measurement. When the heater fork is activated, a net heat current  $\dot {Q}_\mathrm{T}$ is carried through the orifice to the lowermost section. This current is produced by the residual heat leak $\dot{Q}_\mathrm{hl} (\Omega)$ and the heater power $P_\mathrm{h}$. In rotation the current is reduced by the Andreev reflection from vortices $N_\mathrm{eq} (\Omega)$  in the lowermost section.

 The measurement of a carefully prepared reference state with $N_\mathrm{eq} (\Omega)$ vortices is started by demagnetizing the nuclear cooling stage until the thermometer above the lower orifice stops cooling at about $0.20\,T_\mathrm{c}$, \textit{ie.} when its temperature is fixed by the residual heat leak $\dot{Q}_\mathrm{hl}(\Omega)$ and the thermal resistance of the 0.3\,mm aperture. Thereafter the demagnetization cooling is continued at a much reduced rate, to maintain constant conditions. In this situation the temperature below the orifice is much lower, typically $< 0.14 \, T_\mathrm{c}$, as measured in the absence of the 0.3\,mm orifice, owing to good thermal contact via the heat exchange sinter to the nuclear cooling stage. Thus the reverse current of thermal excitations upwards from the lowest section can be neglected and the temperature rise above the orifice depends on the heating delivered by the heater fork and on the rotation velocity, which controls both the Andreev reflection from the vortices below the orifice and the rotation-dependent heat leak to the two compartments above the orifice. By recording the temperature rise as a function of the power $P_\mathrm{h}$ fed to the heater fork, it becomes possible to extract the thermal resistance at the given rotation velocity $\Omega$ and by comparing resistances measured at different $\Omega$, ultimately the Andreev reflection.

The thermometer fork is selected to have small effective mass and a small intrinsic resonance width. In the ballistic temperature regime its damping at low excitation level measures the exponentially vanishing quasiparticle density. Following the analysis of the vibrating wire thermometer, as introduced in Ref.~\cite{WireThermometer}, the resonance width $\Delta f$ of the fork output is approximated as $\Delta f - \Delta f_0 \approx \zeta \exp {(-\Delta/k_\mathrm{B} T)}$, where $\Delta f_0 $ is the intrinsic zero temperature resonance width. In practice, in the regime of linear response with well-behaved Lorentzian resonance line shape it is often sufficient to monitor the resonance amplitude, \textit{i.e.} the in- and out-of-phase signals so that the resonance amplitude and frequency can be recovered, instead of performing complete frequency sweeps across the resonance to measure the width $\Delta f(T)$ directly.

The zero temperature width $\Delta f_0$ is typically measured in vacuum at the lowest possible temperature -- here at about 10\,mK. Its value depends on the type of device, its preparation, mounting, and measuring circuitry, since at best $\Delta f_0 \sim 10$\,mHz which corresponds to a Q value as high as $10^6$ or more. Thus its measurement is a delicate matter, as well as its stability after changes in mounting or simply from one cool down to the next. In Ref.~\cite{MagnonBEC-2} $\Delta f_0$ was determined from in situ measurements of the relaxation rate of trapped magnon condensates as a function of temperature: the result $\Delta f_0 \approx 9\,$mHz was found to agree well with the low-temperature vacuum value of $\Delta f_0$. In Ref.~\cite{ForkThermometer} a tuning fork of similar size and properties was deduced to have $\Delta f_0 \approx 0.13\,$Hz by comparison against a vibrating wire resonator. This value was twice higher than 0.07\,Hz measured in vacuum at 1.5\,K.

For the temperature calibration one needs additionally one reliable temperature reading to fix the second calibration constant $\zeta$. Accurate and at the same time convenient temperature calibration of the fork thermometer is not a straightforward task since a temperature reading from the ballistic regime requires a stable calibration point around $0.2\,T_\mathrm{c}$ or less, where the inevitable temperature difference between an outside thermometer and the $^3$He sample can be large and difficult to estimate. Here $\zeta$ was estimated by extrapolating readings at around $0.3\,T_\mathrm{c}$, by comparison to a $^3$He melting pressure thermometer, which is thermally anchored to the copper nuclear cooling stage, or to B-phase NMR frequency shifts measured in the top $^3$He compartment in Fig.~\ref{Setup} \cite{NMR_shifts}. In Ref.~\cite{ForkThermometer} the calibration was established by comparison to a simultaneously measured vibrating wire resonator.  A 32\,kHz fork with prongs of $0.1 \times 0.22\,$mm$^2$ cross section (where $W=0.1\,$mm is the width perpendicular to the direction of motion) and 2.3\,mm length was concluded to have a calibration constant $\zeta = 43\,$kHz (at zero pressure). One might think that other similar forks would faithfully display the same value, since it is expected to depend only on the density of quartz, on geometrical factors, and the Fermi momentum $p_\mathrm{F}$.

However, the matter appears to be more complicated: Two different forks with similar dimensions and properties as that of Ref.~\cite{ForkThermometer} were used for thermometry in the present rotating measurements with $\zeta = 11.7\,$kHz and 9.4\,kHz at 0.5\,bar pressure, \textit{i.e.} a sensitivity reduced by a factor of 4. In Ref.~\cite{ForkThermometer-2} a larger fork with prongs of $0.35 \times 0.40\,$mm$^2$ cross section and 3.1\,mm length was found to have sensitivity reduced by an order of magnitude from that expected according to the calibration recipe offered in Ref.~\cite{ForkThermometer}. In contrast, the pressure dependence appears to scale $\propto p_\mathrm{F}^4$ as expected, since the fork with $\zeta = 11.7\,$kHz at 0.5\,bar was calibrated at 29\,bar pressure to have $\zeta = 17.5\,$kHz. The ratio of these two values is close to the expected number 1.54.

In Fig.~\ref{TempResponse} we see an example of the temperature response when the excitation of the heater fork is suddenly switched from zero to 6\,pW. The time constant is slow, about 25\,s, determined by the thermal RC time constant of the bolometer, and not by the one order of magnitude faster response which would correspond to the inverse of the resonance width of the thermometer fork. The temperature rise settles at a value of roughly $10\,\mu$K which is small compared to the total temperature difference across the orifice.

\begin{figure}[t!]
\begin{center}
\centerline{\includegraphics[width=0.9\linewidth]{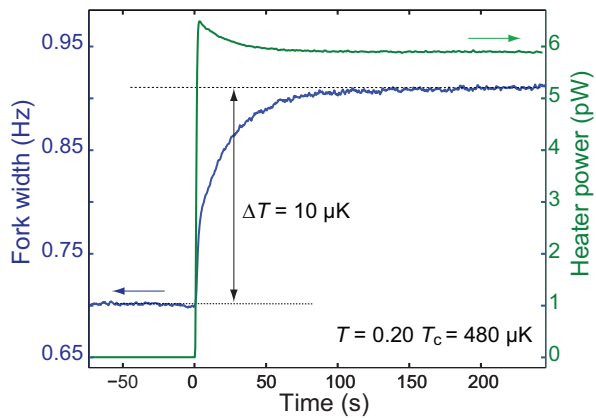}}
\caption{Example of the heater and thermometer responses when the heater is switched on at time $t=0$ and $\Omega = 0$. }
\label{TempResponse}
\end{center}
\vspace{-10mm}
\end{figure}

Assuming thermal equilibrium above the orifice and neglecting the orders of magnitude lower density of quasiparticles below the orifice, one may express the thermal balance across the orifice as
\begin{equation}
\label{balance-1}
\dot{Q}_\mathrm{hl}(\Omega) + P_\mathrm{h} = \dot {Q}_\mathrm{T} \; .
\end{equation}
Here $\dot{Q}_\mathrm{hl}(\Omega)$ is the rotation dependent residual heat leak, introduced by mechanical vibrations and other uncontrolled sources to the two compartments above the orifice. $P_\mathrm{h}$ is the heating power dissipated by the heater fork in the liquid, while $\dot {Q}_\mathrm{T} (\Omega)$ is the heat flow carried by excitations through the orifice, which because of Andreev retroreflection is assumed rotation dependent.

A transparent derivation \cite{BlackBodyRad} of $\dot {Q}_\mathrm{T}$ starts from the kinetic theory of gases where the particle flux from the half space above a pin hole in a thin plate is given by
\begin{equation}
\label{flux}
\Phi_\mathrm{T} = \frac{1} {4} \langle n v_\mathrm{g} \rangle \; .
\end{equation}
Here $\langle n v_\mathrm{g} \rangle = \int \mathcal{N}(E) \, f(E) \, \mathbf{v}_{\rm g}(E) \, dE $ is the thermal average of the excitation density $n(E)$ multiplied by the group velocity $\mathbf{v}_{\rm g}(E)=dE/d\mathbf{p}$ of the excitations, $\mathcal{N} (E)$ is their density of states, and $f(E) \approx e^{-(E/k_\mathrm{B}T)}$ the Fermi distribution function, when $k_B T\ll\Delta$. Thus the particle flux represents an energy flow  $\dot {Q}_\mathrm{T} = \Phi_\mathrm{T}\, \langle E \rangle \, A(\Omega)$, where $ \langle E \rangle = \langle n \, v_\mathrm{g} \, E \rangle / \langle n \, v_\mathrm{g} \rangle \approx \Delta + k_\mathrm{B} T$ and $A(\Omega)$ is the effective area of the orifice. Eq.~(\ref{balance-1}) can now be given in the form
\begin{equation}
\label{balance-2}
\dot{Q}_{hl}(\Omega)+P_{gen}=\frac{4\pi k_B p_F^2}{h^3}Te^{-\frac{\Delta}{k_BT}}(\Delta+k_B T) A(\Omega).
\end{equation}
Andreev reflection is thus measured in terms of the effective orifice area $A(\Omega)$, which is reduced when excitations are  retroreflected. The reduction is conveniently expressed in terms of a reflection coefficient $\nu(\Omega)$, defined as
\begin{equation}
\nu(\Omega)=1-\frac{A(\Omega)}{A(0)}.
\label{ReflCoef}
\end{equation}

\section{Andreev reflection from a vortex array}
\label{Results}

Using Eq.~(\ref{balance-2}), results from measurements as a function of the heating power at different rotation velocities are plotted in Fig.~\ref{ThermalCurrentData}. The data points in this plot are obtained by changing at fixed $\Omega$ the applied power level from one value to the next and by averaging the corresponding equilibrium temperature readings for  $\sim 10$\, min at each power level. The intercept of the linear fit with the power axis gives the residual heat leak $\dot{Q}_\mathrm{hl}(\Omega)$ to the sample volume above the 0.3\,mm orifice. It proves to vary  from 12\,pW at $\Omega = 0$ to 18\,pW at $\Omega = $1.8\,rad/s. It should be pointed out that achieving pW-level heat leaks in a large superfluid $^3$He sample housed within a massive rotating nuclear demagnetization cryostat is a major "tour de force". A large part of the heat leak is caused by residual mechanical vibrations which are enhanced when the rotation velocity is increased, especially on approaching mechanical resonances at certain velocities or owing to the general loss of stability caused by rotational imbalance at high $\Omega$. The $\Omega$ values used in the measurement had to be carefully selected, to be sufficiently spaced from any mechanical resonances. In addition at high $\Omega$ values the rotation-induced heat leak fluctuated with variations of about 1\,pW. Thus it is understandable that both the heat leak and the scatter of the data in Fig.~\ref{ThermalCurrentData} increase with $\Omega$.

\begin{figure}[t!]
\begin{center}
\centerline{\includegraphics[width=0.9\linewidth]{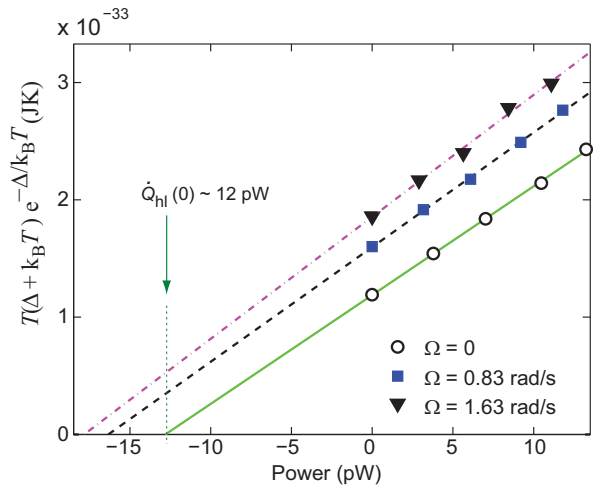}}
\caption{Temperature dependent part of the quasiparticle current in Eq.~(\ref{balance-2}) as a function of the applied heating power at three different rotation velocities. }
\label{ThermalCurrentData}
\end{center}
\vspace{-10mm}
\end{figure}

The inverse of the slope of the lines in Fig.~\ref{ThermalCurrentData} gives the effective area $A(\Omega)$ of the orifice. The measurement with no vortices gives $A(0)\approx0.020$~mm$^2$ which is less than half of the measured geometrical area of the orifice. A hole in an 0.7\,mm thick division plate is not an ideal aperture and diffuse scattering of the excitations from the walls of this channel account for the difference. In any case, it is the relative change in Eq.~(\ref{ReflCoef}) which determines the Andreev reflection which is plotted in terms of the reflection coefficient $\nu(\Omega)$ in Fig.~\ref{ReflCoefVsOmega} as a function of the rotation velocity. The reflection coefficient increases with the total number of vortices $N_\mathrm{v} \propto \Omega$ approximately linearly, which would be the case if the Andreev shadows are additive. Unfortunately, the scatter is large and does not allow resolving whether in the measured arrays the Andreev shadows start to overlap. The main difficulty is not believed to be mechanical vibrations but an unstable heater fork. In principle, if its power calibration would have been time independent, then the calibration would not have affected the measurement of $\nu(\Omega)$.

It is appropriate to remind that the middle compartment above the orifice is also in the equilibrium vortex state. These vortices cause changes in the quasiparticle trajectories above the orifice, but do not give rise to thermal gradients comparable to those created across the orifice, since the orifice is the dominant resistance. Thus sufficient thermal equilibrium is preserved above the orifice also in rotation and the presence or absence of vortices in the upper sections of the tower has little effect on the Andreev measurements.

\begin{figure}[b!!]
\begin{center}
\centerline{\includegraphics[width=0.9\linewidth]{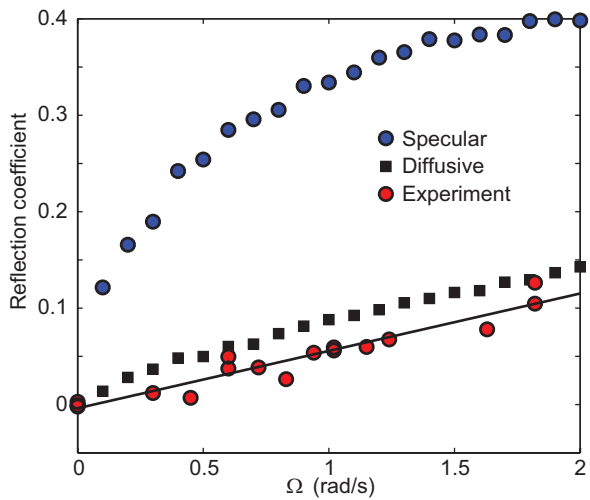}}
\caption{The fraction $\nu(\Omega)$ of quasiparticle excitations Andreev reflected back upward through the orifice, as determined from the measured steady-state temperature increase above the orifice. The temperature is $\sim 0.20\,T_\mathrm{ c}$ above the orifice and$\lesssim 0.14\,T_\mathrm{c}$ below. The statistical uncertainty in the value of $\nu$ is smaller than the size of the data points $(< 2 \cdot 10^{-3})$, but systematic error sources are larger, as discussed in the text. The measured data are in reasonable agreement with simulation calculations applying diffuse scattering from the cylindrical quartz wall. }
\label{ReflCoefVsOmega}
\end{center}
\vspace{-10mm}
\end{figure}

Figs.~\ref{ThermalCurrentData} and \ref{ReflCoefVsOmega} are based on the measured steady state temperature increment when the heater fork is activated. Consistency with the transient response in Fig.~\ref{TempResponse} can be compared by noting that in the thermal time constant $\tau = RC$ the resistance is dominated by the orifice, $ R = [d \dot {Q}_\mathrm{T}/dT]^{-1} \propto 1/A(\Omega)$, and the heat capacity by the specific heat of $^3$He-B in the volume $V$ above the orifice \cite{HeatCapacity},
\begin{equation}
C=k_B \sqrt{2\pi}\mathcal{N}_\mathrm{F}\left(\frac{\Delta}{k_BT}\right)^{\frac{3}{2}} e^{-\frac{\Delta}{k_BT}}\left(\Delta+\frac{21}{16}k_B T\right) V\;,
\end{equation}
where $\mathcal{N}_\mathrm{F}$ is the density of states at the Fermi level. Using the $A(0)$ value from Fig.~\ref{ThermalCurrentData} gives a time constant of 32\,s, which is in reasonable agreement with the 25\,s value in Fig.~\ref{TempResponse}. Thus overall these measurements of Andreev reflection and their analysis are believed to be consistent and to agree with expectations.

The results in Fig.~\ref{ReflCoefVsOmega} have also been checked in numerical calculations with Monte Carlo simulation of individual quasiparticle trajectories. Using a more rigorous formulation, the flux through the orifice is obtained in the form
\begin{equation}
\label{flux}
\dot{Q}_\mathrm{T}(\Omega)=\int \mathcal{N}(E) v_g(E)\, E\,f(E)\C{T}d E d x d y d \varphi d \theta \;,
\end{equation}
where the transmission $\C{T}=\C{T}(E,x,y,\varphi,\theta,\Omega)$ is set equal to one if an excitation above the orifice at position $(x,y)$ and moving in the direction $(\varphi,\theta)$ reaches the sinter and zero if it is Andreev reflected back. The integration is performed over the cross section of the orifice [while $\varphi \in (0,2\pi)$, $\theta \in (0,\pi/2)$  and $E \in (\Delta,\infty)$]. Thus by sampling the trajectories individually by integrating Eq.~(\ref{flux}) numerically and then solving equations (\ref{balance-2}) and (\ref{ReflCoef}) for $\nu$, one collects results for the reflection coefficient.

The simulations were calculated assuming either diffusive or specular quasiparticle scattering from the quartz-glass walls. As seen in Fig.~\ref{ReflCoefVsOmega}, it is the former which are in better agreement with the measurements. In the presence of diffusive scattering a wall collision radically reduces the probability of such trajectories which would reflect the excitation back through the orifice. In contrast, with specular scattering from the quartz walls the reflection coefficient is substantially enhanced in the presence of vortices: an excitation scattering from the wall and subsequently Andreev reflected by a vortex is sent right back through the orifice. The porous sintered heat exchanger surface is assumed to have zero reflectivity.

Overall we might conclude that the calculated data has little scatter, the deviation from linear dependence reflects partial screening of individual Andreev shadows in the vortex array, as analyzed for 2-dimensional random vortex arrays in Ref.~\cite{NumCalc-2}, and the agreement of the measurements with the calculations using diffusive wall scattering and no adjustable fitting parameters is quite satisfactory.

\section{Andreev reflection from moving vortices}
\label{Dynamics}

Andreev reflection leads to variations in the local thermal quasiparticle density if changes occur in the surrounding vortex configuration. Thus  Andreev reflection can be used to monitor vortex motions. A well-known example is the measurement of the free decay of turbulent vortex tangles, which have been created with a vibrating wire \cite{He3Andreev-2} or a vibrating grid resonator \cite{RingDecay,TurbDecay} in a quiescent bath of $^3$He-B. In rotating flow it has been used to record the evolution of well characterized initial states of quantized vorticity in the presence of a time dependent rotation drive.

So far, rotating measurements have concentrated around an elementary question, whether superfluid flow necessarily becomes turbulent in the zero temperature limit, when the bulk mutual friction dissipation $\alpha (T) \rightarrow 0$? This is the general conclusion which one derives from studies of flow in superfluid $^4$He. However, the $^4$He-II vortex core diameter is of atomic size ($\xi \sim 0.1\,$nm) and on this length scale most surfaces tend to be rough so that strong surface pinning can generally be expected to influence dynamic measurements at low velocities and high surface friction to be present at higher velocities.

In $^3$He-B pinning is less conspicuous -- in fact, measurements so far in a cylindrically symmetric fused quartz container, which has been carefully screened with respect to imperfections, have not displayed clear implications from dissipative surface interactions. At high mutual friction dissipation $\alpha > 1$ at temperatures $T > 0.6 \, T_\mathrm{c}$ dynamic responses are laminar and turbulence is observed only at lower temperatures \cite{TurbTransition}. Still, the intriguing possibility remains whether laminar response might be stable in $^3$He-B in the most ideal conditions at zero temperature.

\begin{figure}[t!]
\begin{center}
\centerline{\includegraphics[width=0.98\linewidth]{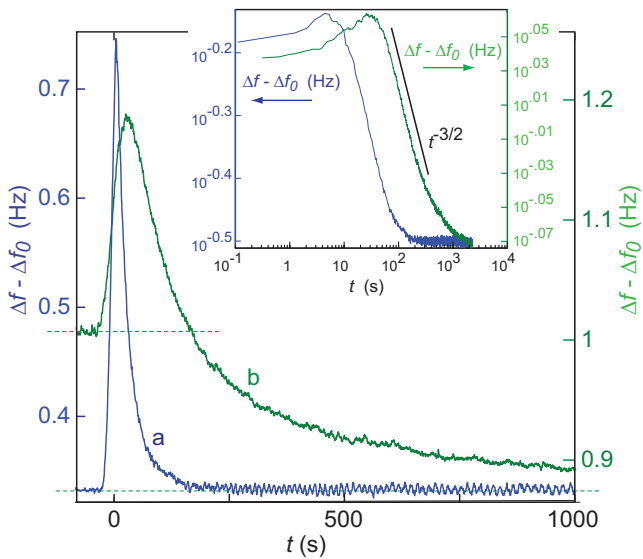}}
\caption{
(Color online) Resonance width of the thermometer fork recorded during free spin-down of the rotating superfluid component. Initially the rotation of the cryostat is brought to rest at a rate $- 0.03\,$rad/s$^2$ and $\Omega =0$ is reached at $t=0$. The rapidly developing large initial peak is generated when the rotating flow is destabilized by turbulence. The turbulent instability appears owing to the breakdown of the rotationally symmetric flow pattern, caused by dissipative interaction with the rough sintered bottom surface. The instability triggers a burst of reconnections, leading to an increase in total vortex length and the formation of a localized tangle near the bottom surface, which together remove about one third of the kinetic energy. The remaining later spin down is predominantly laminar. The two measurements characterize different situations for spin-down from $\Omega_\mathrm{i} = 1.0\,$rad/s: a) 0.5\,bar liquid $^3$He pressure, measuring setup similar to that of Fig.~\ref{AndreevDemo} at $ 0.16\,T_\mathrm{c}$; b) 29\,bar pressure, measuring setup of Fig.~\ref{Setup} where $T \lesssim 0.14\,T_\mathrm{c}$ below the orifice and $0.20\,T_\mathrm{c}$ above. For more explanations about the comparison of these two fork outputs see Sec.~\ref{TurbBurst}.
}
\label{TurbulentSignal}
\end{center}
\vspace{-12mm}
\end{figure}

This hypothesis has been tested in both NMR \cite{LaminarFlowNMR} and Andreev reflection measurements \cite{Hosio3Heflow} with similar conclusions. A straightforward well-defined measurement is obtained by recording the response to a sudden impulsive stop of rotation. Initially the sample cylinder is rotated at constant rotation $\Omega_\mathrm{i}$ in the equilibrium vortex state. Rotation is then suddenly stopped and the free decay of the rotating superflow is monitored while the cylinder itself is at rest, $\Omega =0$. In practice the bulky nuclear demagnetization cryostat of $\sim 300\,$kg cannot be stopped instantaneously without introducing additional heating. Instead, the deceleration to zero rotation is done smoothly at the rate $-0.03\,$rad/s$^2$. The point when the cryostat comes to rest at $\Omega = 0$ is here referred to as $t = 0$. In laminar decay the rotating vortex cluster consists of predominantly straight line vortices which move on a spiral trajectory outward until they annihilate on the cylindrical wall. The outward bound motion is damped only by mutual friction dissipation and therefore the slow laminar response can take hours at the lowest temperatures.

If turbulent processes intervene, then additional dissipation mechanisms are coupled in and the decay speeds up. The motivation driving the research of superfluid dynamics has largely been the question what these mechanisms exactly are. It has been concluded from contactless NMR measurements on the top compartment of the sample cylinder in Fig.~\ref{Setup} that its spin-down response is fully laminar at least down to $0.20\,T_\mathrm{c}$ \cite{LaminarFlowNMR}. It turns out that to obtain turbulence in such a cylinder, the presence of a highly dissipating surface is needed, such as the AB phase boundary separating one section of the cylinder filled with $^3$He-B from another filled with $^3$He-A \cite{AB_Interface}, or a rough surface as in the inset of Fig.~\ref{AndreevDemo} \cite{Hosio3Heflow}, or deviations from cylindrically symmetric flow, such as the obstructions presented by the tuning forks themselves in the compartment in the middle of the cylinder in Fig.~\ref{Setup}.

In such cases of weak turbulent perturbation, laminar flow is initially destabilized at higher vortex density and turbulent disturbances start to evolve: a burst of turbulent tangle formation and of accelerated decay follows, which efficiently reduces the kinetic energy in the early stages of the decay. The turbulent burst is the origin for the initial large peak in the tuning fork output in Fig.~\ref{TurbulentSignal}. This peak is short-lived, and as it subsides, the later response becomes more and more laminar. We describe the laminar signal next and postpone the discussion of the turbulent signal to later.

\begin{figure}[t!]
\begin{center}
\centerline{\includegraphics[width=0.98\linewidth]{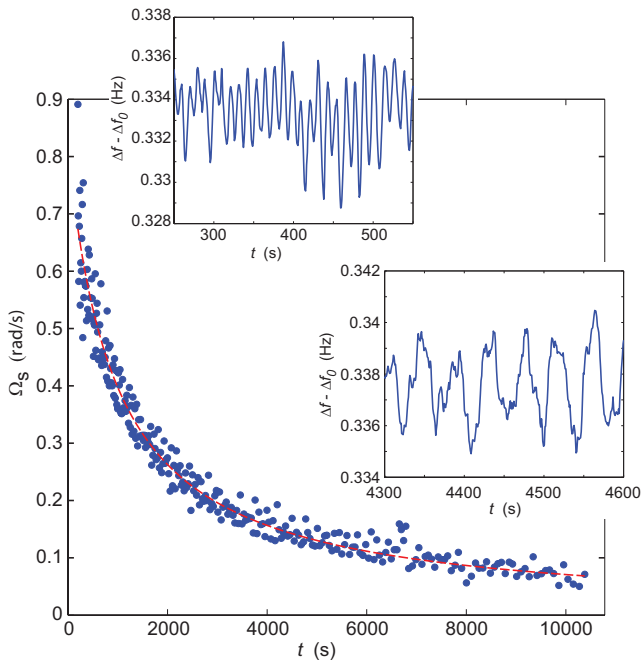}}
\caption{(Color online) Angular velocity $\Omega_\mathrm{s}$ of the superfluid component as a function of time after stopping the rotation of the cryostat from $\Omega_\mathrm{i} = 1.02$\,rad/s. The measurement is fitted to the laminar response in Eq.~(\ref{step}): the dashed line represents $\Omega_\mathrm{s} = 0.815 \, \frac {\mathrm{rad}}{\mathrm{s}} /(1+t/950\,\mathrm{s})$ and gives $\alpha = 6.45 \cdot 10^{-4}$. The two insets show the oscillations of the fork output at the frequency $\Omega_\mathrm{s} (t)$ at two different points in time in the laminar regime of the spin-down decay of Fig.~\ref{TurbulentSignal}. These oscillations of small amplitude arise from the periodic variation in the thermal excitation density around the fork owing to Andreev reflection from the precessing vortex bundle with slight asymmetry. The measurement was performed at 0.5\,bar and $0.16\,T_\mathrm{c}$ in a setup similar to that of Fig.~\ref{AndreevDemo}.
}
\label{LaminarSignal}
\end{center}
\vspace{-12mm}
\end{figure}

\subsection{Laminar response}

A clearly identifiable signature from laminar flow is present in Fig.~\ref{TurbulentSignal} in signal (a). This signal records the resonance width of the thermometer fork as a function of time in a setup similar to that in the inset of Fig.~\ref{AndreevDemo} during the spin-down of the superfluid component. The signal displays a superimposed oscillating component of small amplitude which becomes visible on the falling slope of the rapidly evolving turbulent peak, but grows more prominent after the peak has died at $t > 150\,$s. The oscillations are generated by Andreev reflection from vortices which still are left over and rotate with respect to the stationary cylinder walls. The frequency of the oscillations $\Omega_\mathrm{s} (t)$ monitors the rotation period of the azimuthally circulating flow which is created by this remaining rotating vortex cluster.

An oscillating component arises if the precessing cluster deviates from perfect rotational symmetry, either with respect to its structure or its central confinement within the cylinder, so that Andreev reflection from the periodically changing vortex configuration modulates the thermal excitation density in the neighbourhood of the tuning fork resonator. Although the amplitude of the oscillating signal is small, its presence shows that the rotating bundle of vortices in free spin-down at $\Omega = 0$ does not possess perfect axial symmetry. For instance, a residual misalignment of the rotation and the sample cylinder axes of $\sim 1^\circ $ is expected. In simulation calculations with the vortex filament method \cite{TiltedCylinder} such a small inclination has been concluded to be much below the large tilt angles needed to destabilize laminar flow, so that precession and laminar flow can coexist in the presence of weak breaking of axial symmetry. Oscillating signals from the precession of a vortex bundle with rotational asymmetry have also been observed in NMR experiments \cite{PrecessionNMR,PrecessionFront}.

Analysis of the angular rotation velocity $\Omega_\mathrm{s} (t)$ from the oscillating signal component shows that it follows the laminar decay expected for the vortex density with solid-body density distribution $n_\mathrm{v} = 2\Omega_\mathrm{s} / \kappa$ in mutual-friction-damped spin-down of a vortex cluster composed of straight line vortices. Its time dependence follows from the Euler equation for inviscid rotationally symmetric flow \cite{Sonin}
\begin{equation}
\frac{d \Omega_\mathrm{s}(t)}{dt}=2 \alpha \Omega_\mathrm{s}(t)[\Omega_\mathrm{s}(t)-\Omega(t)].
\label{de}
\end{equation}
Assuming a step change of the rotation drive at $t=0$ from an angular velocity $\Omega_\mathrm{s} (0) = \Omega_0$ to $\Omega =0$, the solution of Eq. (\ref{de}) is given by
\begin{equation}
\Omega_\mathrm{s}(t)=\frac{\Omega_0}{1+t/\tau_\mathrm{L}},
\label{step}
\end{equation}
where the time constant for the decay of laminar flow is $\tau_\mathrm{L}=(2\alpha \Omega_0)^{-1}$. Its value is typically a sizeable fraction of an hour below $0.2\,T_\mathrm{c}$. Thus the laminar decay of rotating superflow conserves the solid-body-like vortex line distribution so that both the density and the total vortex length decay $\propto (1 + t/\tau_\mathrm{L})^{-1}$. In contrast the kinetic energy of the rotating superfluid decays at a steeper rate $\propto (1+ t/\tau_\mathrm{L})^{-3}$, since the flow energy of the rotating superfluid with density $\rho_{\rm s}$ is given by
\begin{equation}
E_{\rm{kin}}=\frac {\pi} {4} \rho_{\rm s}  R^4 h \Omega_\mathrm{s}^2
\label{ekin}
\end{equation}
so that using Eq.~(\ref{step}) the rate can be expressed as
\begin{equation}
\dot{E}_{\rm{kin}}=\frac{\pi} {2} \frac {\rho_{\rm s}  R^4 h \Omega_0^2}{\tau}(1+t/\tau_\mathrm{L})^{-3}.
\label{ekindot}
\end{equation}
Here $R$ and $h$ are the radius and the height of the rotating cylinder.

Returning to our example in Fig.~\ref{TurbulentSignal}, an excellent match between the measured $\Omega_\mathrm{s} (t)$ and Eq.~(\ref{step}) is obtained if one uses $\Omega_0$ and $\tau_\mathrm{L}$ as fitting parameters (Fig.~\ref{LaminarSignal}). Runs at different initial rotation velocities $\Omega_{\rm{i}}=0.6$ --- 1.5\,rad/s and temperatures $T=0.15$ --- $0.19\,T_{\rm c}$ yield rather uniformly $\Omega_0 \sim 0.8\,\Omega_{\rm{i}}$ and as expected $\tau_\mathrm{L} \propto \Omega_0^{-1}$. This means that the relative change from $\Omega_{\rm{i}}$ to $\Omega_0 \sim 0.8\, \Omega_{\rm{i}}$ is not strongly dependent on $\Omega_{\rm{i}}$ or temperature: roughly one fifth of the vortices are removed in the initial turbulent burst, independently of the initial conditions, while the remaining vortices are predominantly straight, oriented along the rotation axis, and decay in a laminar fashion. Their motion towards annihilation at the cylinder wall is resisted by mutual friction only, which here arises from the scattering of thermal excitations from the vortex-core-bound quasiparticles \cite{Kopnin}.

\begin{figure}[t!]
\begin{center}
\centerline{\includegraphics[width=0.8\linewidth]{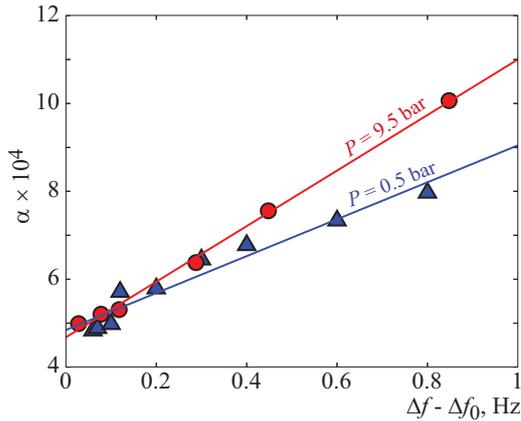}}
\caption{ (Color online) Dissipative mutual-friction parameter $\alpha (T,P)$ in the limit $T \rightarrow 0$, plotted as a function of the measured  resonance width $\Delta f - \Delta f_0 $ of the  thermometer fork in the temperature range 0.14 --- $0.20\,T_\mathrm{c}$. The value of $\alpha$ at 0.5\,bar pressure has been worked out from the precession frequency in the fork width $\Delta f - \Delta f_0$, as shown in Fig.~\ref{LaminarSignal}, by fitting the measured response to Eq.~(\ref{step}), and by averaging over 2 --- 4 measurements at different rotation velocities in the range $\Omega_{\rm{i}}=0.6$ --- 1.5\,rad/s at any given temperature.  The 9.5\,bar data have been obtained from similar recordings of the NMR output. The uncertainty in the $\alpha$ value is estimated to be of order $\pm 5\cdot10^{-5}$. The vertical intercept $\alpha(0)\sim 5 \cdot 10^{-4}$ appears not to vary with liquid $^3$He pressure in the range 0 --- 10\,bar. A finite value of the intercept is a common feature of many similar measurements of ``zero-temperature"  extrapolations involving $^3$He-B vortices, which now needs a solid explanation. }
\label{AndreevAlpha}
\end{center}
\vspace{-12mm}
\end{figure}

In Fig.~\ref{AndreevAlpha} the dissipative mutual friction coefficient $\alpha (T)$, as derived from the fitted results, is shown as a function of temperature on an exponential scale, plotted in terms of the resonance width of the thermometer fork. A linear dependence is obtained, since both mutual friction and the fork width depend on the exponentially temperature dependent excitation density $\propto \exp(-\Delta_0/k_\mathrm{B} T)$. The result is consistent with earlier measurements with a mechanical membrane resonator above $0.35\,T_{\rm c}$ \cite{Bevan} if, in contrast to these authors, no renormalization is applied to the bulk-liquid gap $\Delta (T)$, but instead it is allowed to obey the same temperature dependence as measured by the resonance width $\Delta f (T) - \Delta f_0$ of the fork (see Ref.~\cite{NMR_Alpha} for details). The result is also consistent with NMR data down to $0.20\,T_\mathrm{c}$ at a high pressure of 29\,bar \cite{NMR_Alpha}.

The intriguing feature is the non-zero intercept $\alpha(0)\sim 5 \cdot 10^{-4}$ in the $T\rightarrow0$ limit. The source of this dissipation has not been unequivocally identified, but one explanation involves heating of the vortex-core-bound excitations when vortices are moving with nonzero acceleration, as suggested in Ref.\,\cite{Silaev}. Another possibility could be surface interactions at the rough bottom wall. Other measurements of different type (see eg. Ref.\,\cite{Front}) also point to the existence of new dissipation mechanisms involving vortices, when extrapolated to zero temperature on an exponential scale. The interesting conclusion from Fig.~\ref{AndreevAlpha} is that mutual friction appears to have a lower bound on cooling towards the $T \rightarrow 0$ limit, at least if viewed on an exponentially developing temperature scale. This was not known earlier, but has now important implications on the dynamics of flows in $^3$He-B.

\subsection{Turbulent response}
\label{TurbBurst}

The rapidly evolving initial overshoot in the tuning fork damping in Fig.~\ref{TurbulentSignal} is the signature which we associate with turbulence. The overshoot is an immediate reaction to the deceleration, its maximum is generally reached at the moment when the cryostat rotation $\Omega$ stops, while at high initial vortex density $2 \Omega_\mathrm{i} / \kappa$ the signal may peak already during the deceleration. If the spin-down decay is laminar as in Ref.~\cite{LaminarFlowNMR}, the large but short-lived overshoot is missing, while in the presence of a surface with high vortex friction, like the AB interface in Ref.~\cite{AB_Interface}, a pronounced overshoot in total vortex length and overall dissipation is observed. Thus we expect that the origin for the overshoot of the two signals in Fig.~\ref{TurbulentSignal}, which were recorded  in somewhat different setups, is a turbulent instability caused by interactions of the vortices in the expanding vortex cluster with the rough sintered bottom surface. During rapid vortex decay, when the vortex ends move quickly over the sinter surface, the resulting motion resembles that of uniform surface friction on the AB interface. A second source for instabilities are deviations from cylindrical flow symmetry owing to the presence of tuning forks or their wires. It is the two turbulent responses in Fig.~\ref{TurbulentSignal} which we discuss next.

As the tuning fork monitors the evolution of the local excitation density, other sources than the direct increased dissipation from turbulent vortex decay can also contribute to the signals. For instance, after deceleration the residual heat leak from mechanical vibrations is reduced by $1 \over 3$ because of its $\Omega$-dependence, as seen in Fig.~\ref{ThermalCurrentData}. This reduction is little if at all visible in signal (a), but explains about  half of the drop from the initial to the final signal level in the case of (b). The second half of the overall signal level change in (b) is explained by the Andreev reflection from the initial equilibrium vortex state, which is not present in the final state at $\Omega_\mathrm{s} (t \rightarrow \infty ) = 0$. To create the initial sharp peaks from sources other than the turbulence requires that the deceleration itself generates a pulse of heating. For instance, the rotation of the cryostat displays mechanical resonances with increased vibrational heating at certain fixed rotation velocities. However, these resonances require time to build up which is not available during the rapid deceleration. One can assure oneself about the correctness of such considerations by recording the spin-down signals at different deceleration rates. In short we conclude that the rapidly evolving peaks predominantly characterize the turbulence, but their quantitative analysis is both complicated and different for the two cases in Fig.~\ref{TurbulentSignal}.

If the sinter surface is the main destabilizing effect, then the turbulence is spatially restricted to the vicinity of the bottom surface, as shown by simulation calculations of the free spin down at the AB interface \cite{AB_Interface}. In a long cylinder the volume filled with turbulent tangle and displaying increased dissipation may be assumed to be of similar extension above the AB interface and above a rough sinter surface. However, as the two signals in Fig.~\ref{TurbulentSignal} have been measured in different setups, they display the turbulence differently, primarily since the thermal coupling to the heat sink differs by more than two orders in magnitude. The slower setup (b) was found to have a thermal time constant of 25\,s in Fig.~\ref{TempResponse}. As seen in Fig.~\ref{TurbulentSignal}, the recovery from peak output is not obeying this time constant of the bolometer but takes place on a slower time scale which is determined by the turbulent decay. The bolometric design of setup (b) with the 0.3\,mm diameter orifice emphasizes the increase in fork damping owing to Andreev reflection from the turbulent tangle. Thus this signal is more closely related to the time evolution of the density and polarization of the decaying vortices,  in particular to the total line length deflected towards the plane transverse to the original rotation axis. Since the overall polarization of the vortices changes continuously, while the tangle evolves, the Andreev reflection signal has a complicated origin. In the more open geometry of setup (a) good thermal contact ensures rapid response \cite{Forks-2} and direct heat generation by the turbulence is the dominant source for this output.

Summarizing we note that the short burst of turbulence leads to an accelerated decay of vortices, when compared to the slow laminar decay which dominates the later spin down. The turbulent instability sets in when deceleration is started and causes vortex lines to reconnect and to form a turbulent tangle. This increases the heat release and the total vortex length. Both features are monitored by the fork signals:  (1) The peak in the reconnection rate and dissipation is reached early, already during the deceleration or immediately at its end, depending on the initial rotation $2 \Omega_\mathrm{i} / \kappa$. Subsequently the turbulent dissipation decreases rapidly, as seen from signal (a) in Fig.~\ref{TurbulentSignal}.  (2) The peak in turbulent vortex length develops slower as well as its subsequent decay, as monitored by signal (b). 

The decay of the turbulence proceeds by means of a series of mechanisms. At the outer length scale of the flow disturbance and the turbulent instability, the kinetic energy resides in eddies formed from bundles of approximately aligned vortices. These decay to smaller eddies and bundles by reconnections so that the kinetic energy ends up cascading down the length scales with a Kolmogorov spectrum
\begin{equation}
E(k)=C\epsilon^{2/3} k^{5/3},
\label{kspec}
\end{equation}
where the Kolmogorov constant $C\approx1.5$. Assuming that the kinetic energy is dissipated by some means at the length scale of the inter-vortex distance  $\ell = 1/ \sqrt{L}$, where $L(t)$ is the turbulent vortex density, the dissipation can be assumed on dimensional grounds to be of the form
\begin{equation}
\epsilon = \nu ' \kappa^2 L^2,
\label{eps}
\end{equation}
where $\nu '$ is a phenomenological constant, an effective kinematic viscosity. Equating the cascading energy flux $dE/dt$ from Eq.~(\ref{kspec}) with Eq.~(\ref{eps}) one obtains the time dependence of the turbulent vortex density
\begin{equation}
L (t) = \frac{\sqrt{27C^3}D}{2 \pi\sqrt{\nu '} \kappa} t^{-3/2},
\label{dens}
\end{equation}
where $D$ denotes the long-wave-length cutoff, the length scale of the flow disturbance at the wave vector $k_0 = 2\pi/D$.

These arguments apply to the free decay of a homogeneous and isotropic vortex tangle. It would seem that spin-down in a cylinder with mostly laminar flow of the superfluid component has little to do with full-blown homogeneous turbulence, since reconnections and tangle formation are substantially reduced owing to the high degree of polarization of the vortices along the cylinder axis.  Surprisingly the recovery of both  signals from peak damping in Fig.~\ref{TurbulentSignal} obeys the $t^{-3/2}$ time dependence.

To extract quantitative estimates proves problematic. Depending on whether one assumes the vortices to be mainly parallel or perpendicular to the beam of excitations, one can work out upper or lower bounds on $L(t)$ from signals of type (b), based on numerical calculations of Andreev reflection. This gives a factor of 3 difference in the estimate for the maximum vortex density created in the turbulent burst \cite{Hosio3Heflow}. More sophisticated evaluation of the turbulence in Fig.~\ref{TurbulentSignal} requires comparison to Andreev reflection from numerically calculated simulations of the actual vortex line distribution in the given experimental settings, to which the measurements can be compared.

Moreover, the rate of the decay (Eq.~(\ref{dens})) in terms of one single fitting parameter $\nu^\prime$ has so far not yielded consistent results in the $T \rightarrow 0$ limit: Andreev reflection measurements on the free decay of tangles created with a vibrating grid in $^3$He-B suggest $\nu^\prime  \sim 0.3\,\kappa $ \cite{TurbDecay}, while ion transmission measurements on the free spin down of superfluid $^4$He in a cubic container with rough walls yield $\nu^\prime \sim 0.003\,\kappa$ \cite{Golov}. The magnitude of this disagreement is uncomfortable, but reflects the fact that the measurements and their analyses are not compatible.

To fit the decay rates in Fig.~\ref{TurbulentSignal} requires a further reduction of  $\nu^\prime$ by two orders of magnitude or more,  $\nu^\prime \lesssim 10^{-5} \,\kappa$. It is to be expected that the homogeneous model of Eq.~(\ref{dens}) cannot be appropriate for the weak turbulence with nearly straight vortices in a rotating cylinder with a rough bottom. The analysis of measurements of the opposite case, of how vortices fill a long rotating cylinder at constant $\Omega$, show that the proper description of the precessing and propagating turbulent vortex front, which is then formed, requires two friction coefficients \cite{Front}. The two coefficients are different by two orders of magnitude: a larger bulk friction describes energy dissipation while the smaller accounts for the slow removal of the angular momentum. Thus in the presence of finite vortex polarization, which develops to varying degree in the above mentioned measurements, this decoupling effect, the large difference in decay rates, needs to be taken into account \cite{TurbReview-2}. Clearly, for the signals in Fig.~\ref{TurbulentSignal} these considerations apply: a single energy-dissipation coefficient $\nu^\prime$ is not sufficient to explain the slow rate of angular momentum removal from flow which is only weakly perturbed from cylindrical symmetry.

Considering the above objections, the fact that the two signals in Fig.~\ref{TurbulentSignal} appear to follow the $t^{-3/2}$ time dependence looks to be a coincidence and clearly not a measure of an effective kinematic viscosity. Nevertheless, putting all evidence together, one might argue that a reasonably consistent qualitative interpretation emerges of the interplay of turbulent and laminar flows in $^3$He-B. One particularly pertinent measurement would be a repetition of the experiment in Fig.~\ref{ExpScheme}, but with the cylinder below the orifice replaced by a tube with a square or rectangular circumference and rough walls. In this environment the free spin-down after an impulsive stop of rotation might mimic more closely homogeneous isotropic turbulence and would perhaps provide a more reliable calibration of the Andreev reflection signal from ideal superfluid turbulence, when started from a controlled vortex state. Analogous experiments have been performed with superfluid $^4$He \cite{TurbReview-3}, where turbulent responses dominate. A direct comparison with $^3$He-B would be valuable, in order to compare to a situation when turbulence is less prevalent.

\section{Conclusions}

The zero temperature limit of the Fermi superfluid $^3$He-B, with ballistic quasiparticle transport and no bulk-volume impurity scattering, is in the forefront of current fermion physics in condensed matter, since many new phenomena are here expected to come together. The most recent rush is to search for such alluring predictions as Andreev surface states with excitations of Majorana character. In this effort Andreev reflection is one of the promising experimental tools.

It took a long time before Andreev reflection could be verified in $^3$He-B, but now the practical devices for quasiparticle radiators and sensors have been developed. Further improvement is expected, if arrays of quartz tuning forks \cite{Camera} or of resonators fabricated as micro-electro-mechanical systems (MEMS) become available as quasiparticle detectors. During the past decade it has been demonstrated that the zero temperature limit of $^3$He-B can also be reached and explored in rotation. This advance promises to shed new light on the study of the excitations of the vortex core in this spin triplet orbital p-wave condensate. To search for new evidence one needs extreme low temperatures below $0.15\,T_\mathrm{c}$ and sensitive measuring techniques which function in this temperature range.

So far Andreev reflection in rotation has been employed to calibrate reflection from a well-controlled vortex state and to monitor the free spin-down of the superfluid component in cylindrical flow environments. This calibration measurement is the first and only one where the theory of Andreev reflection from quantized vortices has been quantitatively tested. While the results agree with expectations, more measurements of other well-understood vortex configurations would be valuable.

In NMR measurements the spin-down dynamics have been found to be laminar in ideal cylindrically symmetric flow conditions, which now is understood to result from the decoupling of energy and angular momentum dissipation. Andreev reflection has been applied to measurements where the influence of perturbations is examined, such as the presence of a rough surface which interacts with the end of a moving vortex. The measured responses display an early phase of accelerated turbulent dissipation, with a modest overshoot caused by a burst of reconnecting and tangle formation, and a late phase of slow laminar decay at low vortex density. Unlike in superfluid $^4$He, where the vortex core diameter is of atomic size, in $^3$He-B with two to three orders of magnitude larger core diameters pinning and surface friction are reduced, which helps to reduce dissipation and the duration of turbulence in the dynamic responses. This made it possible to record the angular rotation velocity of a vortex bundle in free spin down, when the damping of a tuning fork resonator is modulated by the variation in the local thermal excitation density, owing to a precessing non-axisymmetric Andreev shadow cast by the bundle. This oscillating signal with slowly increasing periodicity gives directly the laminar vortex friction. It is found to have a small, but nonzero extrapolation at $T = 0$.

This observation has implications on the vortex dynamics of $^3$He-B. For instance, the current view about superfluid turbulence holds that the kinetic energy is transported from large length scales to ever smaller scales owing to a series of different mechanisms. On scales larger than the inter-vortex distance this cascade is structured in a similar manner as in classical turbulence of viscous fluids. On the smaller ``quantum length scales" it is believed that at very low mutual friction dissipation a Kelvin wave cascade becomes possible when nonlinear interactions between Kelvin waves allow the kinetic energy to propagate to smaller wave lengths along a single line vortex.

Such a hypothesis has not been rigorously confirmed experimentally. Numerical simulations at very low mutual friction with sufficient spatial and temporal resolution are notoriously difficult and time consuming. However, recent numerical work \cite{Lvov} concludes that a Kelvin cascade might become possible in the regime $\alpha < 10^{-5}$. Combining this result with Fig.~\ref{AndreevAlpha} means that, unlike in superfluid $^4$He, in the Fermi superfluid $^3$He-B with new sources of dissipation, the Kelvin wave cascade may not be an important component in turbulent dissipation after all. Note, however, that Kelvin wave excitations as such, damped by mutual friction, are most important as a dissipation mechanism. Whether the inference about the absence of the cascade in $^3$He-B holds up in future research will be an interesting question.

Many central questions in vortex dynamics remain unanswered and require further researching. Here rotation and quasiparticle beam techniques, perhaps combined with NMR measurement, will provide new possibilities for noninvasive vortex monitoring.

\vspace{5mm}
\textit{\textbf{Epilogue:}}--Rotating superfluid $^3$He research is deeply indebted to Sasha Andreev who is one of its founding fathers. The first research programme ever to study $^3$He superfluids in rotation was set into motion in 1978 by three academicians Alexander Andreev, Elephter Andronikashvili, and Olli Lounasmaa. This effort, known by the acronym Rota, was supported by both the Academy of Finland and the Soviet Academy of Sciences. It produced the first rotating nuclear demagnetization refrigerator which became operational in mid 1982. Since then the cryostat and its later modifications have been churning out research on quantized vorticity in $^3$He superfluids applying many different measuring techniques. Much of this work has been NMR based and here many generations of students and researchers from the Kapitza Institute for Physical Problems have been participating. Sasha Andreev had the role of an influential godfather, who advanced and promoted the research. In 2012 he was awarded the Lounasmaa Memorial Prize for his achievements in research and its advancement.

\vspace{5mm}
\textit{\textbf{Acknowledgements}}:--We thank our friend and colleague the late Nikolai Kopnin \cite{Kopnin} for promoting and advancing the research on fermion excitations in rotating $^3$He superfluids. This work was financially supported by the Aalto School of Science, its research infrastructure Low Temperature Laboratory, and the Academy of Finland (Centers of Excellence Programme project no. 250280).


\small{$^\ddagger$Present address: Finnish Patent and Registration Office, Arkadia St. 6A, POB 1140, FI-00101 Helsinki, Finland.}

\end{document}